\documentclass{article}
\usepackage[acronym]{glossaries}
\usepackage{authblk}

\newacronym{so}{SO}{SimulatorOrchestrator}

\newacronym{gnn}{GNN}{Graph Neural Network}
\usepackage{cite}
\usepackage[colorinlistoftodos]{todonotes}
\usepackage{amsmath,amssymb,amsfonts}
\usepackage{algpseudocode,algorithm}
\newcommand{\Let}[2]{\State #1$\gets$#2}
\newcommand{\lex}{\ensuremath{\preceq_\textrm{lex}}}
\usepackage{xcolor,booktabs,multirow,subfloat,subcaption}
\usepackage{graphicx}
\usepackage{textcomp}
\usepackage{xcolor}
\usepackage{amsthm}

\def\BibTeX{{\rm B\kern-.05em{\sc i\kern-.025em b}\kern-.08em
    T\kern-.1667em\lower.7ex\hbox{E}\kern-.125emX}}

\usepackage{amsmath}

\begin{document}

\title{AI-Driven Multi-Agent Vehicular Planning for Battery Efficiency and QoS in 6G Smart Cities}

\newcommand{\url}[1]{\texttt{#1}}

\date{}

\author[1]{\small Rohin Gillgallon}
\author[1]{\small Giacomo Bergami}
\author[2]{\small Reham Almutairi}
\author[1]{\small Graham Morgan}

\affil[1]{\footnotesize School of Computing, Newcastle University, Newcastle Upon Tyne, UK}
\affil[2]{\footnotesize Computer Science Department, University of Hafr Albatin, Hafr Albatin, Saudi Arabia}

\maketitle

\begin{abstract}
While simulators exist for vehicular IoT nodes communicating with the Cloud through Edge nodes in a fully-simulated osmotic architecture, they often lack support for dynamic agent planning and optimisation to minimise vehicular battery consumption while ensuring fair communication times. Addressing these challenges requires extending current simulator architectures with AI algorithms for both traffic prediction and dynamic agent planning. This paper presents an extension of \acrfull{so} to meet these requirements. Preliminary results over a realistic urban dataset show that utilising vehicular planning algorithms can lead to improved battery and QoS performance compared with traditional shortest path algorithms. The additional inclusion of desirability areas enabled more ambulances to be routed to their target destinations while utilising less energy to do so, compared to traditional and weighted algorithms without desirability considerations.\\


\end{abstract}

\section{Introduction}
The full adoption of 6G technology implies the integration of AI to enhance the existing 5G infrastructure by supporting decision-making mechanisms. This is deemed necessary when protecting smart city infrastructures, where vehicular traffic incorporates multiple IoT sensors, which can be easy prey and vulnerable to Cybercriminals \cite{10707226,
10770451}. This postulates the adoption of new city simulators with such capabilities.

This paper leverages multi-agent vehicular planning through recent advancements on AI for traffic prediction \cite{DBLP:conf/wimob/MarouaniSJBH24}: when considering IoT vehicles communicating to the Cloud through edges as Road Side Units (RSUs) wihtin a smart city infrastructure,  a large increase in the number of communicating devices threatens to undermine the Quality of Service (QoS), hindering the possibility of sending further information due to bottlenecks. 
Despite 6G cell-free infrastructures paired with the WAN GigaBit token ring network enabling efficient rerouting in city-sized communication areas \cite{s25051591}, 
redirecting vehicular traffic is generally only suitable when we have direct control over a fleet of cars. This has to be done carefully:
we cannot redirect traffic to areas with no communication infrastructure, as this would trivialise the problem, while also considering the desirability of patrolling specific regions. We contextualise this by extending our previous ambulance use case scenario \cite{10770322}: ambulances patrol areas of the city that contain citizens with higher health risks. 
Any routing algorithm should minimise traffic on both road and cyber networks. Achieving this requires solving a multi-objective path optimisation problem. 
As this might easily lead to a factorial number of solutions in the worst case scenario where all possible paths are Pareto optimal, we need to revise state-of-the-art approaches to address real urban communication infrastructures (\S\ref{giacomo} and \ref{sec:potmoast}).

We then extend 
state-of-the-art solutions (\S\ref{reham}) to consider vehicular digital twins reflecting the dynamic fleet of cars over which we have direct control (\S\ref{fleet}) orchestrated by an AI traffic oracle providing feedback to a routing planner (\S\ref{orch}). We propose different algorithms (WSP \S\ref{sec:TDD}, and POTMO-A$\ast$ (\S\ref{sec:potmoast}) for multi-agent vehicular planning, while comparing them to traditional shortest path (SSP) approaches (\S\ref{sec:dspSUMO}) embedded in SUMO. 
Preliminary results over a realistic dataset considering a fleet of ambulances moving around the city while preferring pathways with fewer vehicles and preferring to travel within areas having a higher concentration of frail patients \cite{8280555} shows that {POTMO-A$\ast$ performs the best overall compared to SSP and WSP; resulting in both the best overall network QoS, and ambulances consuming the least energy for all three routes tested; when used within a high desirability area POTMO-A$\ast$ can route more ambulances successfully, whilst using less energy to do so }(\S\ref{rohinResults}). The full configuration for the ambulances being our vehicular digital twin is provided on {\color{red}OSF.io\footnote{\url{https://osf.io/5ykz3/?view\_only=709bd0dad2e94770b4e88815728ba195}}}. {The code extending \gls{so} with vehicular planning is on {\color{red}GitHub\footnote{\url{https://github.com/LogDS/SimulatorOrchestrator/releases/tag/v2.0}}}}. {\color{red}File pointers are given in red.}

\section{Related Works}
\subsection{Simulators for the IoT/Edge/Cloud Continuum}\label{reham}

Simulators targeting the IoT/Edge/Cloud continuum are crucial for evaluating the interaction of distributed orchestration, SDN-based routing, and resource management policies across heterogeneous infrastructures. Several simulation frameworks have been developed to model interactions across the IoT, edge, and cloud layers \cite{s24051511}. IoTSim‑Edge extends the popular CloudSim platform by modelling heterogeneous edge devices, battery usage, communication protocols, and mobility within realistic applications \cite{jha2020iotsim}. It facilitates the evaluation of energy consumption and QoS under diverse configurations. 
EdgeCloudSim~\cite{edgecloudsim} provides abstractions for latency modelling and service placement across fog and cloud layers, while lacking support for mobile IoT devices or 6G networks. 

IoTSim-Osmosis-RES~\cite{szydlo2022iotsim} introduces renewable energy-aware orchestration and autonomic scheduling. Nevertheless, it does not consider fine-grained vehicle behaviour or dynamic network state updates influenced by real-world mobility conditions. While SimulatorBridger \cite{simulatorbridger} monitors the energy efficiency of electric vehicles within realistic traffic simulations, it does not simulate network-layer communication or QoS-sensitive packet transmissions across edge/cloud environments.

\gls{so}~\cite{s25051591} was initially developed to support osmotic communication for static patient digital twins: we extend the former to orchestrate fleets of ambulances navigating an urban setting. As this is mediated through SUMO, we also integrate vehicle mobility. In contrast, the previous \gls{so} architecture enables vehicular IoT devices at the edge, as determined by RSUs in SUMO. Communication events are dynamically routed through SDN-controlled osmotic infrastructure in \gls{so}. The system generates real-time feedback from traffic-aware edge nodes, enabling ambulances not only to avoid congested areas but also to prioritise zones with high concentrations of elderly or frail populations. 
Each simulation timestep involves transforming mobility events into communication sessions, modeled at the Session and Presentation layers of the OSI stack. Simulating all seven layers was deemed impractical due to the computational cost and the complexity of simulating heterogeneous IoT protocols. While full-stack security analysis is beyond the current scope, prior work has shown that OSI-layer-specific vulnerabilities—particularly in VANETs—can affect both energy consumption and QoS~\cite{10707226}. 

While simulators are starting to incorporate 6G-oriented features -- such as ultra-low latency modeling, distributed AI orchestration, edge-native service placement, and flexible architectural models~\cite{6G} -- they typically remain domain-specific. \gls{so} addresses this gap by providing unified support for SDN-based orchestration, agent-based vehicular mobility, and osmotic edge-cloud communication. Its architecture enables the coordination across mobility, communication, and computation layers, for real-time agent planning, battery minimisation, and QoS enforcement in 6G-aware smart city scenarios.

\subsection{Real-Time AI Agent Planning through Graph Search}\label{giacomo}
\textbf{Time Dependent Dijkstra (TDD)} \cite{DBLP:journals/ior/Dreyfus69} extends Dijkstra's algorithm with edges' weights being a function of time. 
\textbf{Lifelong Planning A*}  \cite{NIPS2001_a5910243} extends traditional A* algorithms by considering a map for locally preserving the Bellman equations for undiscounted deterministic sequential decision problems \cite{Bel}. Similarly to A*, the $g$ cost can be used to reconstruct the path from the goal to the start of the search. 
These solutions do not consider the multidimensional cost of the edges.

\textbf{Multiobjective Dijkstra} \cite{MARISTANYDELASCASAS2021105424} generalises Dijkstra's, as each edge comes with a $d$-dimensional cost in  $\mathbb{R}_{\geq0}^d$. Pareto optimality is employed as a search heuristic to efficiently derive the shortest path. This algorithm considers neither time-dependent edge weights nor provides the possibility of having multi-dimensional edge costs that change over time. \textbf{MOPBD*} \cite{DBLP:journals/ral/RenRLC22} generalises Lifelong Planning A* by considering multi-objective optimization. It returns all possible paths between the source and goal states with Pareto-efficient costs. In the worst-case scenario, this might lead to dealing with a considerable number of solutions if all these paths between source and target are Pareto optimal: given a complete graph $G=(V,E)$ containing at least two vertices, the number of all possible simple paths between two nodes is $\nu=\lfloor (|V|-2)! e\rfloor$. 

This paper considers a priority ordering over each specific dimension, thereby attempting to prioritise the scores by considering them in lexicographical order. By retraining only the locally Pareto optimal solutions per node, while considering only one possible path per graph according to the order above, we drastically reduce the total number of potential solutions to one, thereby reducing the time complexity.



\section{Extensions to \acrfull{so}}\label{simorch}

{
\subsection{AI for Traffic Prediction and Fleet Orchestration}\label{orch}
The simulator aims to support 6G AI algorithms by extending its capabilities in two ways. \textit{First}, we extend the simulator to get live information concerning how many vehicles are communicating with each of the edge nodes within the environment, as well as {the number of active communications $\#e$ on a road $e$ among others. 
This extension allows for vehicles to be dynamically re-routed using real-time information without relying on historical data (\S\ref{sec:TDD}). 
\textit{Second},  a traffic prediction oracle forecasts future traffic information patterns based on local information enables multi-agent planning over historical data (\S\ref{sec:potmoast}). After collecting historical data from the number of devices communicating through \gls{so}, we train a GNN to predict traffic trends (\S\ref{sec:trafficpred}), which then serves to determine dynamic traffic edge weights.
Whilst \cite{s25051591} did allow for the injection of dynamic vehicular information, this simulation did not consider dynamic rewriting for a specific class of vehicles. We enable this by re-adapting the static digital twin technology to be used as electric vehicles moving within the SUMO simulation. The communicating vehicle's position is then transferred to SO.}

\subsection{Electric Vehicle Digital Twins as Dynamic Agents}\label{fleet}
Within the previous implementation of \gls{so} \cite{s25051591,icit} the traffic and IoT simulators were run sequentially. The traffic simulator, SUMO in this case, would generate a full set of location-timestep pairs. 
After this, the IoT simulator would use these to determine at which times a vehicle would have generated a communication towards the cloud via the edge and then establish such a communication. 
By now, using SUMO to route vehicles within a realistic traffic scenario, the ambulances' live positions in SUMO are fed to \gls{so} at each simulation time in a structure containing the ID, position, vehicle type, and SUMO simulation time.
All previous models of IoT-Osmosis-RES models considered neither consumption strategies nor actual transmission bitrates specific to the particular IoT device; our current solutions take this into account through the energy model available in Mininet-WiFi \cite{7367387}. As such, we assume that all cars are equipped with Bit ZigBee transmitters  \cite{1440950,1019408}, addressing the needs of low-cost and low-power wireless IoT data networks. 

\subsection{Extension to the Packet Routing algorithm} \label{PRA}
We improve SPMB Quietest from \cite{s25051591}, supporting normal vehicular connection counting to update the MEL values before allocating MIPs to the edge nodes. Now, if there are multiple 'best' MELs, select the geographically closest 'best' MEL instead of the first alphabetically, improving realism while ensuring to keep the communications as near as possible to the area starting it. By doing so, we should strengthen the correlation between the number of vehicles in a given area and the area of started communications.

\section{Vehicular Digital Twin Planning}\label{shortestpath}
The vehicular agent orchestration problem can be solved by running several times a priority-ordered, timed, and multi-objective shortest path calculation problem. 

\textbf{Problem Formulation:} Given a finite temporal multigraph $G=(V,E,c_t,h_t)$, where $V$ (and $E\subseteq V\times V$) is a finite set of vertices (and edges) and $c_t\colon E\to \mathbb{R}^d$ (and $h_t\colon E\to \mathbb{R}^d$) is the multidimensional (heuristic) cost of traversing an edge $e\in E$ at time $t$ where a dimension $d_t\in \mathbb{N}_{\neq 0}$ refers to the estimated time required to traverse the edge, we want to find a simple path between $s$ and $t$ starting at time $\tau_0$ minimising the overall traversal cost while considering cost priorities over the lexicographic order over the $d$-dimensional vectors $\preceq_\textup{leq}$. Given $\Pi_{s,t}$ the set of all the possible simple paths connecting $s$ with $t$, we therefore want to determine the following minimization:
\begin{equation}
    {\min\arg}_{\pi\in\Pi_{s,t}}\sum_{e_i\in \pi}F(\pi,i)
\end{equation}
{where $F$ is the recurrent relation $F(\pi,i)=c_{\tau_0+F(i-1)[d_t]}(e_i)$ for $|\pi|\geq i\geq 1$ and $e_i\in \pi$, and as $0$ otherwise.} 

For testing the possibility of using historical traffic data for forecasting traffic congestion for the vehicular route planner, we first train a traffic predictor (\S\ref{sec:trafficpred}) to then use the predicted weight values as one of our dimensions for multidimensional path costs (\S\ref{sec:potmoast}). We compare this strategy against exploiting real-time dynamic information (\S\ref{sec:TDD}) while keeping  Dijkstra algorithm as a baseline (\S\ref{sec:dspSUMO}).

\subsection{Traffic Prediction using \gls{gnn}}\label{sec:trafficpred}
\gls{gnn}s \cite{47094} provide a specialised neural network architecture considering graphs as inputs. Through the usage of pairwise message passing, graph nodes continuously update their representations by exchanging information with their neighbours: this feature makes it extremely relevant to simulate traffic predictions, as neighbouring nodes can exchange information concerning current traffic loads in specific points of the city. As an input to the \gls{gnn}, we consider a fully connected graph, where each node represents an RSU providing traffic information as collected by \gls{so}. In contrast, each edge $e$ represents the shortest path between two RSUs. We consider unidimensional features for both nodes and edges: while for nodes, we provide the time-varying traffic volume, we consider as many edge features $\textup{exp}(f(e))$ as the possible information $f(e)$ that we can extract from one single path: the shortest distance between the two points (MinLen), the time it takes to traverse the path when the way is free (MinTime), and the number of edges to be traversed (Hop). Within this framework, each node in the graph performs univariate time series forecasting regarding the vehicular distributions of the node over time. 

\subsection{Ambulance Routing Strategies}\label{ars}
To motivate the need for our proposed algorithm, we provide intermediate traffic orchestration solutions, remarking the inadequacy of single-cost functions in determining this.

\subsubsection{SUMO's Shortest Path (SSP)}\label{sec:dspSUMO} {{\color{red}SUMO\footnote{\url{https://sumo.dlr.de/docs/Simulation/Routing.html}}} uses Dijkstra algorithm \cite{Dijkstra} to dynamically route vehicles by mainly considering distance metrics. This is our baseline.}

\subsubsection{Weighted Shortest Path (WSP)}\label{sec:TDD} {the {\color{red}TraCI Python library\footnote{\url{https://sumo.dlr.de/docs/TraCI.html}}} enables dynamic edge update costs, where each edge $e$ represents a road between two junctions. SUMO will then recalculate the best SSP path from the current vehicle position towards destination $t$ for a specific vehicle $\alpha$. As \gls{so} updates the communicating IoT devices at each simulation tick, we  re-weight the edge cost for each ambulance using the live \gls{so}  $\#e$ information (see \S\ref{orch}). 
We can then recompute the edge weight per ambulance as:
\begin{equation}
    \phi(e,\alpha) = \dfrac{\|\varpi_e- \varpi_t\|_2}{\max(\|\varpi_\alpha- \varpi_t\|_2,\varepsilon)} \cdot \left(\|\varpi_e- \varpi_t\|_2 + w(\alpha)\right) \cdot \#e
\end{equation} 
where $0<\varepsilon\ll 1$, and $w(\alpha)$ is the time $\alpha$ has been stationary in a traffic jam and $\varpi_x$ denotes SUMO's projection of a physical point $x$ into a Cartesian coordinate system with distance expressed in meters. The weights, $\phi(e,\alpha)$, are calculated for each road into/out from each edge node ($e$) at each simulation step. Algorithm \ref{Algorithm:2} illustrates the aforementioned data processing.

\begin{algorithm}[H]
	\caption{Weighted SUMO Routing (WSP)} 
	\begin{algorithmic}[1]
       \Procedure {updateEdgeWeightes}{ambuID, SOinfo, edgeNodesPos, DestPos}
            \ForAll{edges $\in$ edgeNodes}
            \State edgeNodeDistance = distance(edgeNodePos.getPos(edgeNode), destPos) 
                \ForAll{road $\in$ \Call{getEdgeNodeIncomingEdges}{edgeNode}}
                    \State road.Weight = edgeNodeDistance + getWaitingTime(ambuID)
                    \State ambuDistFromStart = getDistanceFromStart(ambuID) 
                    \State road.Weight = roadWeight $\cdot$ edgeNodeDistance / distance(edgeNodePos.getPos(edgeNode) , ambuStartPos) $\cdot$ SOInfo
                    \State rerouteVehicle(ambuID)
                \EndFor
            \EndFor
        \EndProcedure
	\end{algorithmic}
 \label{Algorithm:2}
\end{algorithm}

\subsubsection{POTMO-A$\ast$} \label{sec:potmoast}
we run the time-varying and multi-objective shortest path for each agent after determining the best path for each. By considering one dimension as the number of predicted vehicles traversing the edge, we can gradually increment this as soon as we know that one ambulance will traverse the edge. These considerations enable the solution to the problem as outlined in the Problem Formulation. 

As a multi-objective optimization problem cannot guarantee that a single solution optimizes each objective, we then decide to relax the problem to allow for a hierarchy of costs to be considered in different order of priority as follows: (\textit{i}) the number of predicted communicating cars occurring within the area within the slot of interest from historical data, (\textit{ii}) expected battery consumption given actual road slope and length over the vehicular Digital Twin battery model, (\textit{iii}) area desirability value, (\textit{iv})  haul time given the maximum velocity speed on the current road, and (\textit{v}) actual length of the road segment. We derive heuristic costs by revising the former dimensions after considering a beeline from any point of interest towards the target destination 
at maximum 50km/h speed. 
To fill in the gaps of the car distributions considering only values near osmotic Edge nodes, we apply the Savitzky–Golay filter \cite{sgfilter}.

\begin{algorithm}[!p]
	\caption{Priority-Ordered Timed Multi-Objective  A$\ast$
		\label{alg:packed-dna-hamming}}
	\begin{algorithmic}[1]
		
		\algrenewcommand\algorithmicindent{0.5em}%
		\Require{$\lex$: $n$-dimensional vector lexicographical ordering}
		\Require{$P$: Pareto front for some $n$-dimensional vectors.}
		\Require{$MC$: some $n$-dimensional costs with weighted paths}
		\Require{$d_t$: dimension of time in $n$-dimensional costs.}
		\Require{$\oplus$: increasing \textit{lhs} MC costs by  \textit{rhs} value and extending textit{lhs} MC paths with \textit{rhs} edge}
		\Statex
		\Function{POTMO-A$\ast$}{$G=(V,E,c_t,h_t),s,\tau_0,t$}
		\Let{$F$}{\textbf{new} PriorityQueue($[(\tau_0,s),\vec{0}]$)}\label{line:pq}
		\Let{Cost}{\textbf{new} Map()} \Comment{Cost$\colon V\mapsto MC$}
		\Let{OpenSet}{$\{s\}$}
		\Let{first}{\textbf{true}}
		\While{$F\neq\emptyset$}
		\Let{curr,$\tau'$}{\Call{Pop}{$F$}}
		\State \algorithmicif\; curr=$t$\;\algorithmicthen\;\Return Cost(curr)
		\Let{OpenSet}{OpenSet$\backslash\{\textrm{curr}\}$}
		\Let{OldCost}{Cost(curr)}
		\ForAll{$e\in E_\textrm{out}(curr)$ \textbf{s.t} \texttt{dst}($e$)$\neq$curr}
		\If{\textbf{not} first \textbf{and} \texttt{dst($e$)$\in$OldCost}}\label{line:ignore}
		\State \textbf{continue}
		\EndIf
		\Let{$w$}{$c_{\tau'}(e)$}
		\Let{$\delta,\tau''$}{$w[d_t],\tau'+w[d_t]$}\label{line:arrival}
		\Let{$nc$}{Cost(curr)$\oplus(w,e)$}
		\If{$\texttt{dst}(e)\notin \textup{dom}$(Cost)}
		\Let{Cost$(\texttt{dst}(e))$}{$nc$}\label{line:firstenc}
		\State \Call{Push}{$F$, $[(\tau'',\texttt{
				dst}(e)),\min_{\lex} nc+\delta+h_{\tau'}(e)]$}\label{line:inprio1}
		\Let{OpenSet}{OpenSet$\cup\{\texttt{dst}(e)\}$}
		\Else
		\Let{$oc$}{Cost$(\texttt{dst}(e))$}
		\State\algorithmicif\,$P(oc)=oc$\,\algorithmicthen\,\textbf{continue}\label{line:noimprove}
		\Let{Cost$(\texttt{dst}(e))$}{$P(oc)$}\label{line:setPareto}
		\State\algorithmicif\, $\texttt{dst}(e)\in$OpenSet \,\algorithmicthen\,\textbf{continue}\label{line:inOpenSet}
		\State \Call{Push}{$F$, $[(\tau'',\texttt{
				dst}(e)),\min_{\lex} nc+\delta+h_{\tau'}(e)]$}\label{line:inprio2}
		\Let{OpenSet}{OpenSet$\cup\{\texttt{dst}(e)\}$}
		\EndIf
		\EndFor
		\EndWhile
		\EndFunction
	\end{algorithmic}
\end{algorithm}

We discuss how it Algorithm \ref{alg:packed-dna-hamming} 
works by marking the main differences with classical A$\ast$. 
By considering a priority queue considering not only the nodes to be currently considered but also the time when such nodes are supposed to be reached given the car digital twin model of interest: 
this is because we might expect to reach the same node with different costs depending on the time we are expected to arrive there. We originally initialise such a queue with a zero multidimensional vector as a cost for reaching the current source $s$ by starting the navigation at time $\tau_0$. We ignore considering neighbouring nodes for which we obtain no substantial cost change from previous runs; 
we also ignore further navigating the edge if the Pareto Front obtainable from navigating the current node does not improve over the previously estimated costs 
and if the node is already scheduled to be visited as part of the OpenSet. 
Given the function $c_{\tau'}(e)$ returning the multidimensional cost $w$ for traversing edge $e$ at time $\tau'$, by extracting the estimated navigation time $\delta$ we derive the expected arrival time $\tau''$. 
The cost to reach the target is calculated as the Pareto front of solutions leading to the target, 
which boils down to the currently considered cost if this is the first time we encounter the target node. 
After increasing each cost in $nc$ for reaching the target is increased by the time required to traverse the edge $\delta$ and the heuristic value to reach the destination ($h_{\tau'}(e)$), we consider the costs being Pareto optimal and appearing first after lexicographic order $\preceq_{\textup{lex}}$. Such a value is then added to the priority queue. 

\section{Empirical Evaluation}
We consider non-agentic traffic patterns for the SUMO Bologna ringway \cite{7234936} {containing 22\,213 unique vehicles}. Each RSU is an osmotic 6G-Edge node through which any IoT vehicle, both from SUMO and orchestrated as \S\ref{shortestpath}, can communicate with the cloud through \gls{so}.  
 We derive the desirability values from an elderly patient frail distribution \cite{8280555}.  

\subsection{GNN Traffic Prediction}

\begin{table}[!b]
\centering
\caption{GNN-based traffic prediction for our scenario.}\label{differentDatasets}
\resizebox{.7\columnwidth}{!}{\begin{tabular}{ lccc }
\toprule 
\multirow{2}{*}{Scores} & \multicolumn{3}{c}{Bologna} \\\cmidrule{2-4} 

& MinLen & MinTime & Hop \\ \midrule 

DTW  &  $\color{blue}3.57 \times 10^3$ & $3.71 \times 10^3$ & $\color{red}3.79 \times 10^5$ \\ 
WDDTW  &  $\color{blue}1.11 \times 10^5$ & $1.17 \times 10^5$ & $1.17 \times 10^5$ \\ \midrule 
MAE & \textcolor{blue}{0.61} & 0.63 & \textcolor{red}{0.67} \\ 
MRSE & 1.25 & \textcolor{red}{1.27} & \textcolor{blue}{1.20} \\ 
 \bottomrule
\end{tabular}}
\end{table}


We implemented our GNN using PyTorch and the MultiLayer wrapper, considering three foundational networks: one for \textit{node} features, a second for \textit{edge} features, and a \textit{global} model that updates the predicted graph with features trained from the former two layers. Each of these three consists of a MultiLayer Perceptron with 5 layers having the following activation functions: Linear, Unary Batch Normalization, ReLU, and a Linear model. The neuron size of their hidden layers depends on the foundational network: 64 for {node}} and global and 128 for the {edge} features. {\color{red}See \texttt{gnn.py} on our GitHub.}


We trained the \gls{gnn} to minimise the mean absolute error (MAE), while the root mean squared error (RMSE) is for testing, as MAE gave better stability and faster convergence in our preliminary results.  As these give no information on the distance between expected and predicted time series per node, we also test the predictions by computing both Dynamic Time Warping (DTW) \cite{10.1145/3230734}  and Weighted Derivative Dynamic Time Warping (WDDTW) \cite{JEONG20112231}. \tablename~\ref{differentDatasets} shows that most of the provided metrics agree and consider MinLen the best hyperparameter for this dataset. We conduct our QoS and battery experiments while retaining this configuration. 





\subsection{Testing QoS and Battery Consumption}\label{rohinResults}

\begin{figure}[!h]
    \centering
    \includegraphics[width=\textwidth]{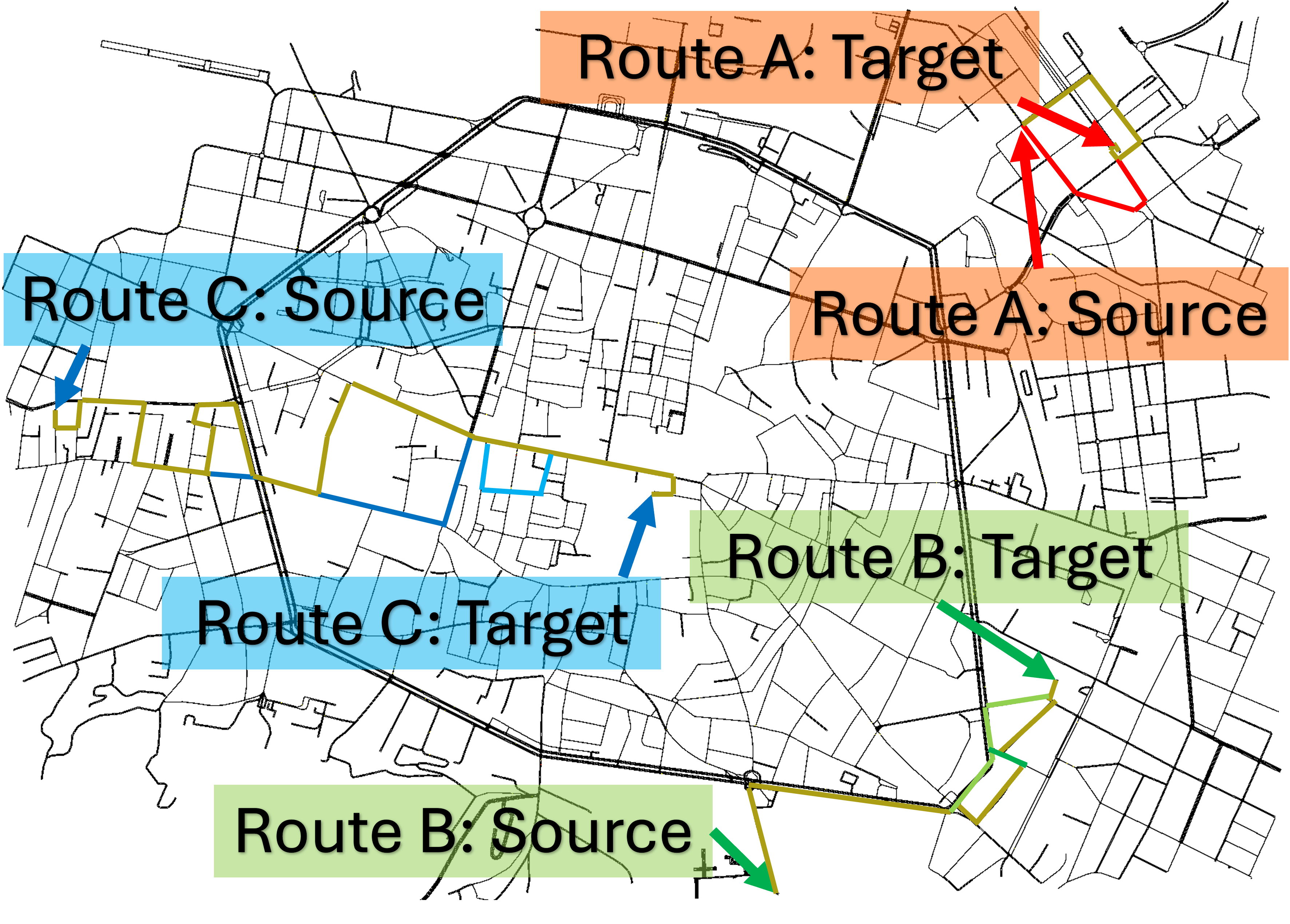}
    \caption{Extending the Bologna dataset \cite{7234936} with ambulances' source and target points for 3 routes of interest (A,B,C). The gold lines are POTMO-A$\ast$ routes, while the rest refer to routes generated by SSP or WSP.}\label{mortadella}
\end{figure}

Our experiments extend the SUMO dataset by considering a fleet of {35} electric ambulances simulated within our vehicle digital twin; this reflects a realistic number of ambulances for the given city. These are tested along the routes in \figurename~\ref{mortadella}.
{{To accurately determine the efficacy of the routes from \S\ref{ars}, the ambulances need to navigate an environment with realistic traffic information. As such, trace data for the other vehicles needs to be used despite the connection counting assumption with \gls{so}, see \S\ref{PRA}. SPMB is still used within \gls{so} due to the connection counting data, simulating the city orchestrator not necessarily having knowledge of any IoT device IDs other than those of the ambulances We also assume that a single ambulance is introduced every 100 seconds{, while all vehicles' have a packet transmission rate of 60 seconds.}} As the former set of experiments remarked that GNN trained with the MinLen edge feature lead to the best results, we consider as the values for the first dimension for POTMO-A$\ast$ as the ones predicted from this model. As the results indicated, none of the configurations led to zero classification errors, thus not overfitting the data. Thus, any prediction resulting from the oracle might reflect traffic trends, but not necessarily provide the exact traffic information as provided by the model. Given this, any advantage coming from the POTMO-A$\ast$ over WSP should be ascribable to the application of the Savitzky-Golav filter smoothing out the traffic distribution to the neighbouring nodes, as well as considering each cost component distinctly, despite not reflecting real-time information. Thus, the following experiments aim to test the benefit of considering a multi-objective optimisation problem rather than weighting it, despite the use of real-time information from the simulator.

\setlength{\textfloatsep}{0pt}
\begin{table}[!t]
\centering
\caption{Comparing the ambulances successfully reaching the destination (ARD) and the total combined energy consumed by the ambulances (TEC) for each of the three routing techniques for each of the three routes.}
\label{tab:1}
\resizebox{\columnwidth}{!}{\begin{tabular}{ lcccccc }
\toprule 
\multirow{2}{*}{Route} & \multicolumn{2}{c}{SSP} & \multicolumn{2}{c}{WSP}& \multicolumn{2}{c}{POTMO-A$\ast$} \\ \cmidrule{2-7}
& ARD & TEC (Wh) & ARD & TEC (Wh) & ARD & TEC (Wh) \\ \midrule
A & {\color{blue}34} & $\color{red}5.39\times 10^6$& {\color{blue}34} & $5.07\times 10^6$ & {\color{blue}34} & $\color{blue}3.95\times 10^6$ \\ \midrule
B & {\color{blue}30} & $\color{blue}2.03\times 10^7$ & 23 & $\color{red}2.46\times 10^7$ & {\color{red}17} & $\color{blue}2.03\times 10^7$ \\ \midrule
C & {\color{red}19} & $2.27\times 10^7$ & 20 & $\color{red}2.60\times 10^7$& {\color{blue}23} & $\color{blue}1.57\times 10^7$ \\
 \bottomrule
\end{tabular}}
\end{table}

\begin{figure*}[!t]
    \begin{subfigure}[b]{0.5\textwidth}
        \includegraphics[width=\textwidth]{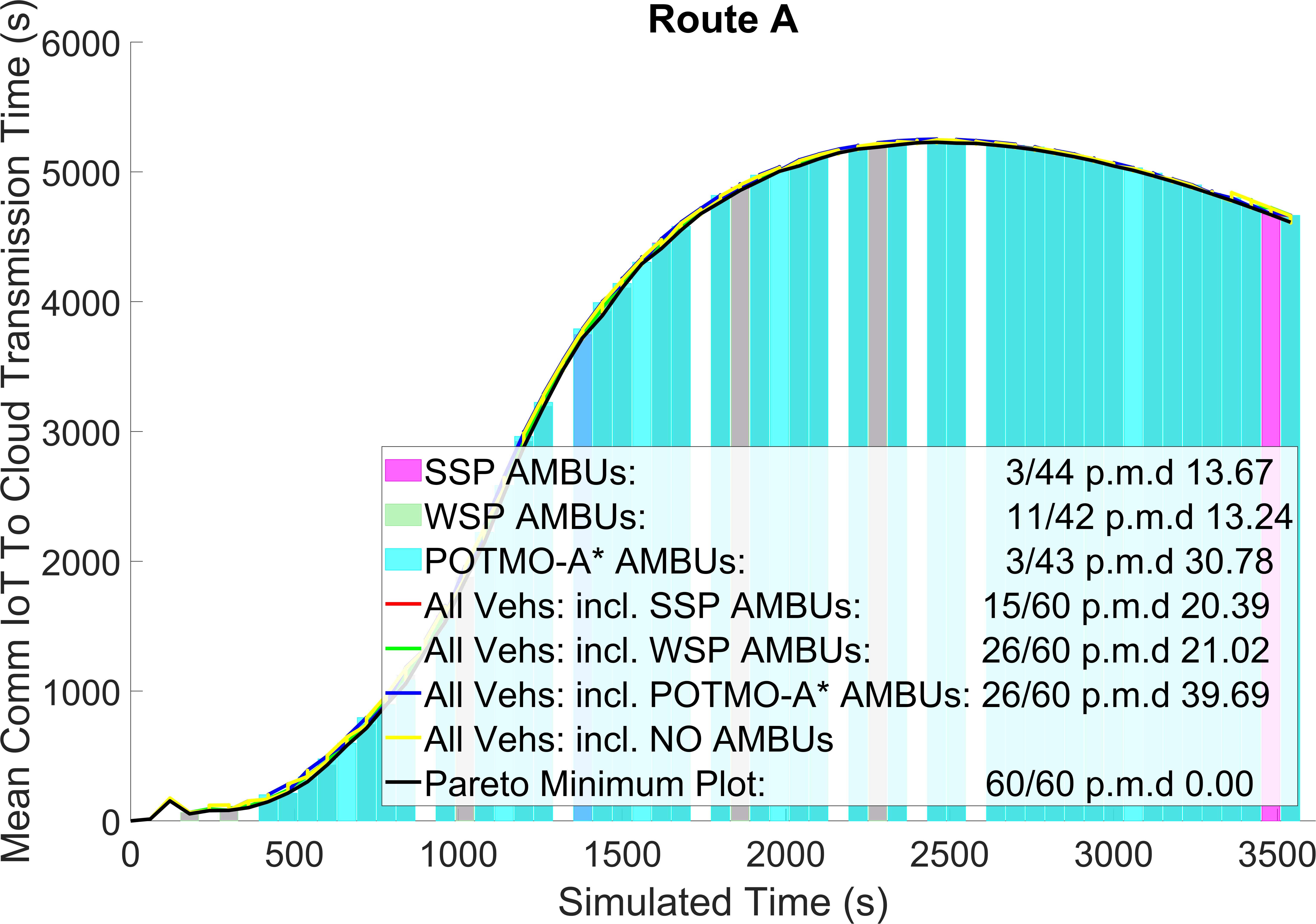}
        \caption{Route A}
        \label{fig:1a}
    \end{subfigure}
    \begin{subfigure}[b]{0.5\textwidth}
        \includegraphics[width=\textwidth]{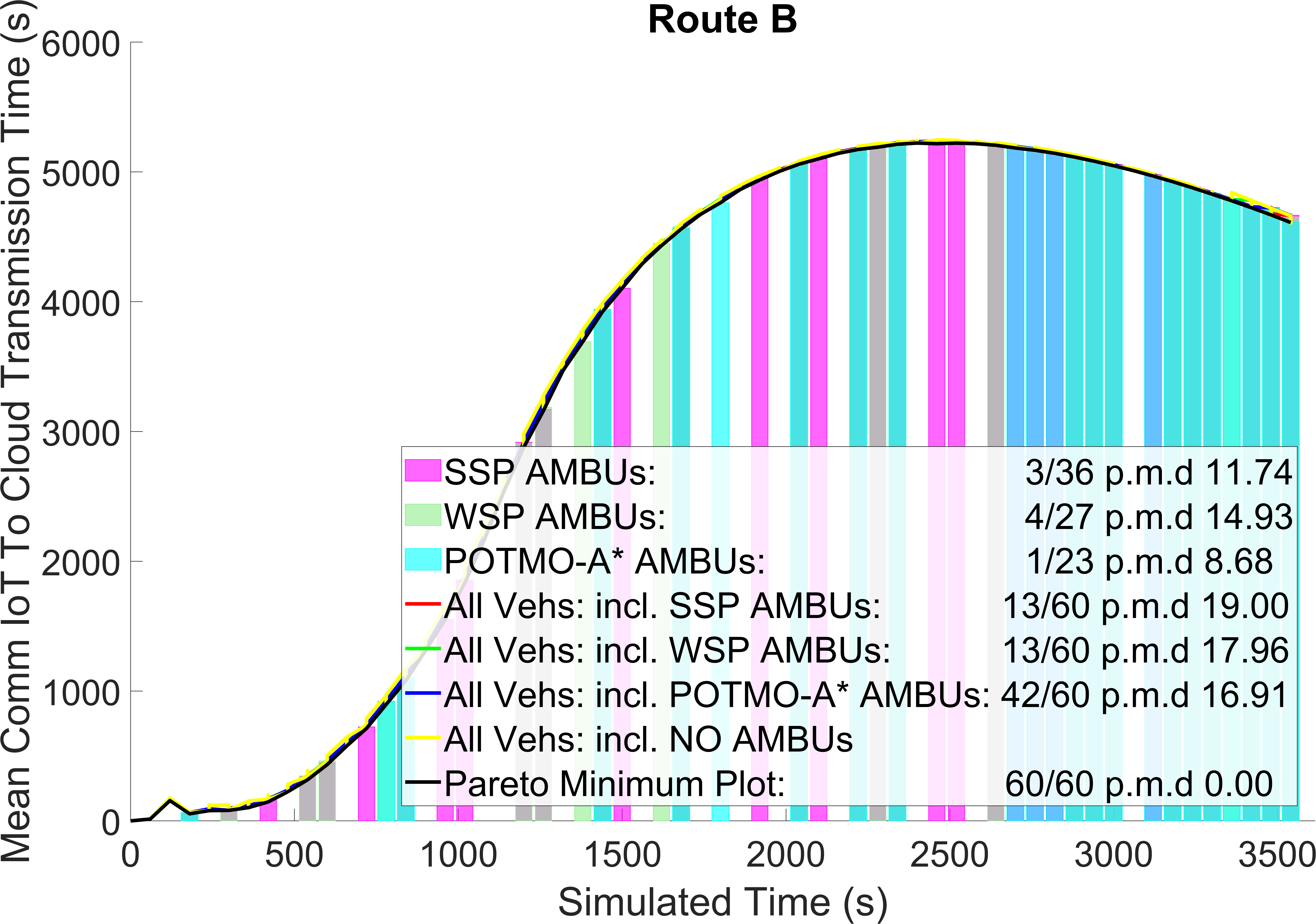}
        \caption{Route B}
        \label{fig:1b}
    \end{subfigure}
    \begin{subfigure}[b]{\textwidth}
        \centering
        \includegraphics[width=0.5\textwidth]{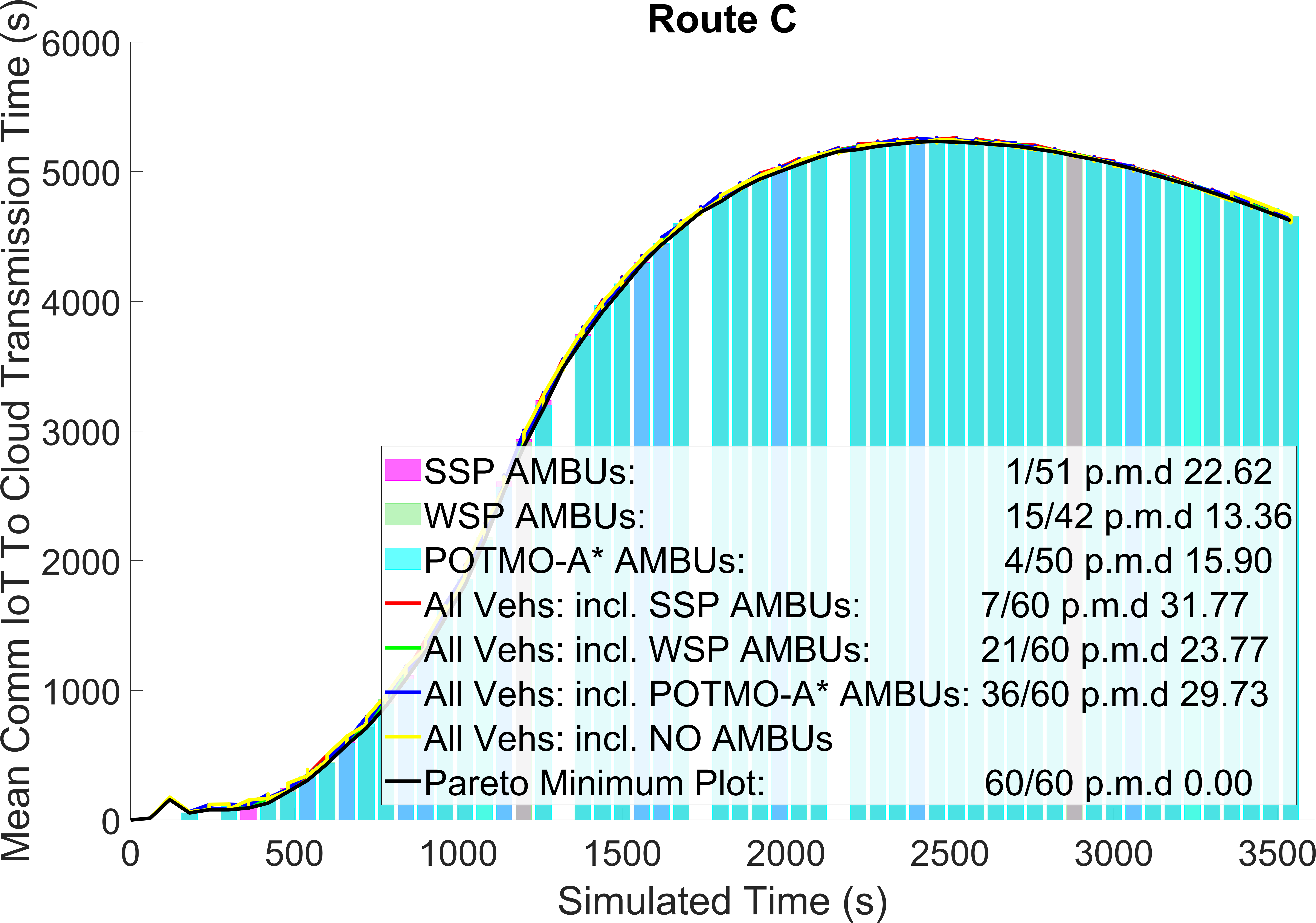}
        \caption{Route C}
        \label{fig:1c}
    \end{subfigure}
    \caption{{The bar plots show the mean transmission times for communications from the ambulances to the cloud; line plots show the transmission times for all communications from all vehicles for each of the three routing techniques for each of the three routes as well as a run with no ambulances; along with a Pareto minimum plot (BLACK line), the fractions in the legend show how many time steps from each run are equal to the Pareto minimum, along with the Pareto Mean Difference (p.m.d.) being the overall mean difference from all time steps from the Pareto minimum. GREY bars are overlaps of SSP with WSP.}}
    \label{fig:1abc}
\end{figure*}
\subsubsection{Battery Consumption} \label{battcons}
{The first test of the routing techniques was concerning the energy efficiency of the generated routes. From Table \ref{tab:1}, each routing technique resulted in a different amount of energy consumed by the ambulances to reach the same destination. Whilst for all 3 routes POTMO-A$\ast$ resulted in the lowest combined total energy consumed (TEC) for all 35 ambulances, it can only be considered optimal for routes A and C. For these routes POTMO-A$\ast$ not only has the lowest TEC but the, at least joined, highest ARD, meaning the POTMO-A$\ast$ routes are the most efficient and the most resilient to changing traffic conditions. For Route B, SSP is the optimal algorithm, with joined best TEC and best SSP. {The likely explanation as to why POTMO-A$\ast$ performs so poorly, routing less than half of the 35 ambulances to their destination, for Route B is region desirability. The region desirability aims to keep ambulances where the most frail patients are, and most of them are within the centre of the city, which in turn explains why POTMO-A$\ast$ was best for Route C. However, there are very few frail patients in the bottom right region and so POTMO-A$\ast$ performs less well. To address this shortcoming, ambulances could switch to SSP when not within the city centre, or a different algorithm could be trained with high desirability for rural regions, and ambulances could choose between POTMO-A$\ast$ and this new algorithm depending on their destination. Future works will also extend POTMO-A$\ast$ to consider additional weights improving over these results.}

WSP, despite using live information, was only able to completely outperform SSP for Route A, resulting in the same ARD with a lower TEC; for Route C WSP resulted in a higher ARD, but also a higher TEC, which given these are ambulances may be preferred. For Route B, it was worse for both TEC and ARD. This is likely due to \gls{so} only having vehicular information concerning junctions with communicating nodes: as no filter is applied to guess vehicular distributions in other regions, weights for junctions with an edge node and a few vehicles may be higher than junctions with no edge node and many vehicles. This can be overcome by only increasing the number of communicating nodes.}

\subsubsection{Transmission Time (QoS)} 
{Given the route taken dictates which edge nodes an ambulance drives past and when, ambulances will be communicating with the network at different times depending on their given route; this experiment tests both the resilience of our packet routing algorithm (\S\ref{PRA}) as well as the adequacy in maintaining QoS through multi-agent vehicular planning. \figurename~\ref{fig:1abc} provides little variations: the overall communication times appears very similar to the Pareto Front across all different ambulance scenarios. The `NO AMBUs' runs (yellow line, not considered in the Pareto Front) refer to runs of the simulator over the same scenarios but without the ambulances. From \figurename~\ref{fig:1abc} the route taken by the ambulances has only a slight impact on the network QoS, as 'NO AMBUs' closely follows the other plots. However, as the ambulances transmit and receive patient data in potentially time-critical situations, every time gain is of the utmost importance. {Each fraction in \figurename~\ref{fig:1abc} remarks how many times a solution lies on the Pareto Front: when considering all combined vehicles POTMO-A${\ast}$ is optimal for each of the 3 routes, tied with WSP for Route B. Although given the time-critical communications are specifically to and from the ambulances, `WSP AMBUs' in fact appears best for all 3 routes. However, as mentioned in \S\ref{battcons}, the current implementation of WSP essentially diverts ambulances away from edge nodes with many surrounding vehicles, which means if an ambulance is near to an edge node it is more likely to be quiet - the drawback as discussed in \S\ref{battcons} is ambulances may be instead directed to more busy areas.} Regardless, looking at the Pareto Mean Difference (p.m.d.) values (the mean deviation from Pareto Front) the highest p.m.d. across all 3 routes is 39.69s, just 0.98\% of  4055.06s - the mean 'IoT-to-cloud-communication-transmission-time' for all communications across all 3 routes.}

\section{Future Works}

{Despite POTMO-A$\ast$ re-routing ambulances to near where fragile people live, it does not support the scheduling of rescue activities, as it would require further recalculating the path to collect a patient and then quickly moving back to the hospital with the patient in need.} 
Future works will  consider potential extensions of our AI-driven routing algorithms to also consider these in a live Smart City scenario \cite{Grenouilleau_Hoeve_Hooker_2019}. 
Future works will also consider longer routes requiring battery recharge by also considering the cost of fully recharging the battery or quantising it by representing battery recharging stations as specific nodes alongside the time necessary for the recharge.  
This will also require extending the vehicular digital twin with 
battery recharge model like PyChargeModel \cite{osti_1883311}. 

\end{document}